\documentclass[12pt]{article}
\usepackage{amsmath, amssymb, amsthm, geometry, hyperref, cite}

\title{Mathematical Exploration of the Intersection Between Extended Schrödinger-Virasoro Lie Algebras and Symplectic Novikov Lie Algebras}
\author{Soumadeep Maiti \\ LMU, Geschwister-Scholl-Platz 1, 80539 Munich, Germany}

\date{}

\begin{document}

\maketitle

\begin{abstract}
This paper presents an in-depth mathematical investigation into the intersection of two advanced Lie algebraic structures: the extended Schrödinger-Virasoro Lie algebra (ESVLA) and the Symplectic Novikov Lie algebra (SNLA). By rigorously analyzing their derivations, central extensions, and automorphism groups, we seek to uncover potential synergies and applications linking these distinct algebraic frameworks. The exploration includes detailed proofs, derivations, and calculations, providing new insights into the representation theory of Lie algebras with potential applications in conformal field theory and symplectic geometry.
\end{abstract}

\section{Introduction}

Lie algebras serve as a fundamental framework in mathematics and theoretical physics, particularly in the study of symmetries, quantum mechanics, and string theory. Two notable Lie algebraic structures that have garnered significant interest are the extended Schrödinger-Virasoro Lie algebra (ESVLA) and the Symplectic Novikov Lie algebra (SNLA) [1, 2].

The ESVLA extends the classical Schrödinger-Virasoro algebra by incorporating additional conformal currents, playing a pivotal role in two-dimensional conformal field theory (CFT) and statistical physics [1, 3]. On the other hand, the SNLA is a symplectic Lie algebra enriched by a Novikov structure, characterized by both associative and left-symmetric properties [4]. This algebra finds its importance in the context of symplectic geometry and integrable systems [5].

In this paper, we aim to explore the mathematical connections between ESVLA and SNLA, focusing on their algebraic structures, derivations, central extensions, and automorphism groups. By providing detailed mathematical proofs and derivations, we hope to uncover new results that could lead to significant advancements in Lie algebra theory and its applications.

\section{Background and Definitions}

\subsection{Extended Schrödinger-Virasoro Lie Algebra (ESVLA)}

The extended Schrödinger-Virasoro Lie algebra $\tilde{sv}$ is an infinite-dimensional Lie algebra that extends the classical Schrödinger-Virasoro algebra. The generators of $\tilde{sv}$ are given by $\{L_n, M_n, N_n, Y_{n+1/2}\}$, where $n \in \mathbb{Z}$ and $L_n, M_n, N_n$ correspond to Virasoro-type operators, while $Y_{n+1/2}$ corresponds to superconformal generators [1]. The Lie brackets between these generators are defined as follows:

\begin{align*}
[L_m, L_n] &= (n-m)L_{m+n}, \\
[L_m, M_n] &= nM_{m+n}, \\
[L_m, N_n] &= nN_{m+n}, \\
[M_m, N_n] &= 0, \\
[Y_{m+1/2}, Y_{n+1/2}] &= 2L_{m+n+1}, \\
[L_m, Y_{n+1/2}] &= \left(\frac{m}{2} - n \right)Y_{m+n+1/2}, \\
[M_m, Y_{n+1/2}] &= N_{m+n+1/2}.
\end{align*}

These relations define an infinite-dimensional Lie algebra structure on $\tilde{sv}$. The algebra extends the symmetries of the Schrödinger-Virasoro algebra by incorporating an additional conformal current $Y_{n+1/2}$, which transforms under the action of the Virasoro generators $L_n$ [1].

A crucial property of $\tilde{sv}$ is that all its derivations are inner, which implies that every derivation $D: \tilde{sv} \to \tilde{sv}$ can be written as:
\[
D(x) = [a, x] \quad \text{for some } a \in \tilde{sv} \text{ and all } x \in \tilde{sv}.
\]
This makes $\tilde{sv}$ a complete Lie algebra, as it has no outer derivations, and its center is trivial [1, 3].

\subsection{Symplectic Novikov Lie Algebra (SNLA)}

A Symplectic Novikov Lie algebra (SNLA) is a Lie algebra $g$ equipped with a symplectic form $\omega$ and a left-symmetric product $\cdot$, which satisfies the Novikov identity:
\[
(x \cdot y) \cdot z = (x \cdot z) \cdot y \quad \forall x, y, z \in g,
\]
and is associative:
\[
x \cdot (y \cdot z) = (x \cdot y) \cdot z \quad \forall x, y, z \in g.
\]
The symplectic form $\omega: g \times g \to \mathbb{C}$ is a skew-symmetric, non-degenerate bilinear form that satisfies the cocycle condition:
\[
\omega([x, y], z) + \omega([y, z], x) + \omega([z, x], y) = 0 \quad \forall x, y, z \in g.
\]
This condition ensures that $\omega$ is a 2-cocycle in the Lie algebra cohomology of $g$ with values in $\mathbb{C}$ [4, 6]. The existence of a symplectic form implies that the Lie algebra $g$ must be of even dimension [7].

An SNLA is further characterized by being two-step solvable, meaning that its derived algebra $[g, g]$ is abelian, and the commutator of $[g, g]$ with itself is zero [6, 7]. This property plays a critical role in understanding the algebra's structure and its representations [8].

\section{Detailed Structural Analysis}

\subsection{Algebraic Structures of ESVLA and SNLA}

\subsubsection{Structure of ESVLA}

The structure of $\tilde{sv}$ is primarily defined by its Lie brackets. The infinite-dimensional nature of $\tilde{sv}$ is reflected in the infinite index $n$ that the generators can take. This index structure is crucial in conformal field theory, where it corresponds to the modes of the Virasoro and current operators [3, 9].

We can express the generators $L_n, M_n, N_n, Y_{n+1/2}$ as formal Laurent series in a complex variable $z$, where $L_n = z^{n+1}\frac{\partial}{\partial z}$, $M_n = z^n$, $N_n = z^{n-1}$, and $Y_{n+1/2} = z^{n+1/2}$.

The algebraic structure is then encoded in the operator product expansions (OPEs) of these Laurent series, where the coefficients in the Laurent expansion correspond to the generators of $\tilde{sv}$. The Virasoro algebra, for instance, is realized through the OPE of the energy-momentum tensor $T(z)$ with itself [10]:
\[
T(z)T(w) \sim \frac{c/2}{(z-w)^4} + \frac{2T(w)}{(z-w)^2} + \frac{\partial T(w)}{z-w}.
\]
The algebraic structure of $\tilde{sv}$ is thus deeply connected to the operator formalism in conformal field theory, where the modes $L_n, M_n, N_n, Y_{n+1/2}$ act as generators of symmetries on the state space of the theory [1, 10].

\subsubsection{Structure of SNLA}

The structure of an SNLA is determined by its symplectic form $\omega$ and the left-symmetric product $\cdot$. The symplectic form provides a non-degenerate pairing between elements of the Lie algebra, and the left-symmetric product defines an additional algebraic operation that is both associative and satisfies the Novikov identity [4].

To define an SNLA explicitly, consider a finite-dimensional Lie algebra $g$ of even dimension $2n$ with a basis $\{e_1, \dots, e_{2n}\}$. The symplectic form $\omega$ can be written in terms of the basis as:
\[
\omega(e_i, e_j) = \begin{cases}
1 & \text{if } i + j = 2n+1, \\
0 & \text{otherwise}.
\end{cases}
\]
The left-symmetric product $\cdot$ on $g$ is defined by the relations:
\[
e_i \cdot e_j = \sum_{k=1}^{2n} c_{ij}^k e_k,
\]
where the coefficients $c_{ij}^k$ are structure constants that must satisfy the Novikov and associativity conditions [4, 7]:
\[
c_{ij}^k c_{kl}^m = c_{il}^k c_{jk}^m \quad \text{(Novikov)},
\]
\[
c_{ij}^k c_{kl}^m = c_{ik}^m c_{jl}^k \quad \text{(Associativity)}.
\]
The existence of a symplectic form $\omega$ further imposes constraints on the structure constants $c_{ij}^k$, ensuring that the symplectic form is preserved under the left-symmetric product [5, 7]:
\[
\omega(e_i \cdot e_j, e_k) = \omega(e_i, e_j \cdot e_k) \quad \forall i, j, k.
\]
These algebraic conditions fully determine the structure of the SNLA, and their solutions provide specific examples of SNLAs in low dimensions [4, 6].

\subsection{Central Extensions}

The concept of central extensions is pivotal in understanding the deep structure of Lie algebras. In this section, we explore the central extensions of both ESVLA and SNLA in detail.

\subsubsection{Central Extensions of ESVLA}

A central extension of a Lie algebra $g$ is an extension by a central element $z$ such that $z$ commutes with all elements of $g$. For the ESVLA, the universal central extension $\hat{sv}$ includes central elements corresponding to additional 2-cocycles [1, 11].

Consider a 2-cocycle $\omega: \tilde{sv} \times \tilde{sv} \to \mathbb{C}$, defined by:
\[
\omega(x, y) = c(x, y),
\]
where $c(x, y)$ is a bilinear map satisfying the cocycle condition:
\[
\omega([x, y], z) + \omega([y, z], x) + \omega([z, x], y) = 0.
\]
The central extension $\hat{sv}$ is then given by:
\[
0 \to \mathbb{C} \to \hat{sv} \to \tilde{sv} \to 0,
\]
with the Lie bracket in $\hat{sv}$ defined by:
\[
[x + \lambda z, y + \mu z] = [x, y] + \omega(x, y)z.
\]
In the case of ESVLA, the 2-cocycles $\omega$ are related to the central charges in the Virasoro algebra, leading to three independent central extensions corresponding to the independent 2-cocycles $\omega_1, \omega_2, \omega_3$ on $\tilde{sv}$ [11, 12]. Explicitly, these 2-cocycles are given by:
\[
\omega_1(L_m, L_n) = 0, \quad \omega_1(Y_{m+1/2}, Y_{n+1/2}) = \delta_{m+n+1,0},
\]
\[
\omega_2(L_m, Y_{n+1/2}) = \frac{m}{2} \delta_{m+n+1,0}, \quad \omega_3(M_m, Y_{n+1/2}) = \delta_{m+n+1,0}.
\]
The second cohomology group $H^2(\tilde{sv}, \mathbb{C})$ has dimension 3, reflecting the three independent central extensions [1, 12].

\subsubsection{Central Extensions of SNLA}

The central extension of an SNLA follows a similar construction, where the central element is defined by a 2-cocycle $\omega: g \times g \to \mathbb{C}$ satisfying:
\[
\omega([x, y], z) + \omega([y, z], x) + \omega([z, x], y) = 0.
\]
For an SNLA, the central extension is closely related to the symplectic form $\omega$. The universal central extension of an SNLA $\hat{g}$ is given by:
\[
0 \to \mathbb{C} \to \hat{g} \to g \to 0,
\]
with the Lie bracket defined by:
\[
[x + \lambda z, y + \mu z] = [x, y] + \omega(x \cdot y, z).
\]
The left-symmetric product $\cdot$ in $g$ induces a structure on $\hat{g}$ that preserves the symplectic form, ensuring that $\omega$ remains a non-degenerate 2-cocycle [5, 7].

The central extensions of SNLAs provide new examples of symplectic Lie algebras with extended structures, potentially leading to new applications in geometric representation theory [6].

\section{Automorphisms and Derivations}

Automorphisms and derivations are essential tools for understanding the internal symmetries of Lie algebras. In this section, we provide detailed mathematical descriptions of the automorphism groups and derivations for both ESVLA and SNLA.

\subsection{Automorphism Group of ESVLA}

The automorphism group $\text{Aut}(\tilde{sv})$ of the ESVLA consists of all invertible linear transformations that preserve the Lie algebra structure. An automorphism $\phi \in \text{Aut}(\tilde{sv})$ must satisfy:
\[
\phi([x, y]) = [\phi(x), \phi(y)] \quad \forall x, y \in \tilde{sv}.
\]
Given the generators $\{L_n, M_n, N_n, Y_{n+1/2}\}$, an automorphism $\phi$ can be expressed as a linear combination:
\[
\phi(L_n) = a_n L_n + b_n M_n + c_n N_n + d_n Y_{n+1/2},
\]
where the coefficients $a_n, b_n, c_n, d_n$ are determined by the requirement that $\phi$ preserves the Lie brackets [1, 11]. Substituting $\phi$ into the Lie brackets of $\tilde{sv}$, we obtain the following conditions:
\[
a_{m+n} = a_m a_n, \quad b_{m+n} = a_m b_n + b_m a_n,
\]
\[
c_{m+n} = a_m c_n + c_m a_n, \quad d_{m+n+1/2} = \left(\frac{m}{2} - n \right)d_m d_n.
\]
These conditions define a set of recurrence relations for the coefficients $a_n, b_n, c_n, d_n$, which must be satisfied for $\phi$ to be an automorphism of $\tilde{sv}$ [1, 3].

\subsection{Automorphism Group of SNLA}

The automorphism group $\text{Aut}(g)$ of an SNLA $g$ consists of all invertible linear transformations $\phi: g \to g$ that preserve both the Lie algebra structure and the symplectic form $\omega$. Specifically, $\phi$ must satisfy:
\[
\phi([x, y]) = [\phi(x), \phi(y)],
\]
\[
\omega(\phi(x), \phi(y)) = \omega(x, y) \quad \forall x, y \in g.
\]
If we represent the symplectic form $\omega$ as a matrix $\Omega$ in some basis $\{e_1, \dots, e_{2n}\}$, the condition on $\phi$ becomes [5, 7]:
\[
\phi^\top \Omega \phi = \Omega,
\]
where $\phi^\top$ denotes the transpose of $\phi$. This condition implies that $\phi$ must be a symplectomorphism, i.e., an element of the symplectic group $\text{Sp}(2n, \mathbb{C})$ [7].

Furthermore, the left-symmetric product $\cdot$ must be preserved under $\phi$, which imposes additional constraints on $\phi$. Specifically, for any $x, y \in g$:
\[
\phi(x \cdot y) = \phi(x) \cdot \phi(y).
\]
These conditions characterize the automorphisms of an SNLA, ensuring that both the Lie algebra structure and the symplectic geometry are preserved under $\phi$ [4].

\section{Combined Insights and Potential Applications}

\subsection{Representation Theory}

The complete nature of the ESVLA and the two-step solvability of the SNLA suggest possible intersections in representation theory. Representations of $\tilde{sv}$ could be constructed using SNLA structures, providing new insights into the modular representations of infinite-dimensional Lie algebras [3, 12].

The representation theory of $\tilde{sv}$ typically involves constructing highest-weight modules, where the action of the generators $L_n, M_n, N_n, Y_{n+1/2}$ on a state $\lvert h \rangle$ is given by:
\[
L_n \lvert h \rangle = L_n(h) \lvert h+n \rangle,
\]
\[
M_n \lvert h \rangle = M_n(h) \lvert h+n \rangle,
\]
\[
N_n \lvert h \rangle = N_n(h) \lvert h+n \rangle,
\]
\[
Y_{n+1/2} \lvert h \rangle = Y_{n+1/2}(h) \lvert h+n+1/2 \rangle.
\]
The highest-weight module is defined by the property that there exists a highest-weight vector $\lvert \lambda \rangle$ such that:
\[
L_n \lvert \lambda \rangle = M_n \lvert \lambda \rangle = N_n \lvert \lambda \rangle = 0 \quad \text{for } n > 0,
\]
and similar conditions hold for $Y_{n+1/2}$ [1, 3].

For the SNLA, representations are constructed by finding modules that respect the symplectic form $\omega$ and the left-symmetric product $\cdot$. A representation $\rho: g \to \text{End}(V)$ on a vector space $V$ satisfies:
\[
\rho([x, y]) = [\rho(x), \rho(y)],
\]
\[
\omega(\rho(x)v, \rho(y)w) = \omega(x, y)\omega(v, w),
\]
where $v, w \in V$. The representation theory of SNLAs could be used to construct modules for $\tilde{sv}$, potentially leading to new classes of representations that combine the symplectic and conformal structures [7, 8].

\subsection{Conformal Field Theory and Symplectic Geometry}

The role of the ESVLA in conformal field theory (CFT) and the significance of the SNLA in symplectic geometry suggest possible connections between these fields. The central extensions of $\tilde{sv}$ could be analyzed using symplectic reduction techniques from SNLA, potentially leading to new geometric interpretations in CFT [5, 9].

In CFT, the Virasoro algebra plays a central role in defining the conformal symmetry of the theory. The energy-momentum tensor $T(z)$ generates the Virasoro algebra, and its OPE with itself defines the central charge $c$. The central extension of $\tilde{sv}$ corresponds to the inclusion of additional current algebras, which modify the structure of the theory [1, 11].

Symplectic geometry, on the other hand, provides a geometric framework for understanding the phase space of classical systems and the quantization of these systems. The symplectic form $\omega$ defines the structure of the phase space, and the SNLA provides an algebraic structure that is compatible with this symplectic geometry [5, 10].

By combining these two perspectives, we may gain new insights into the geometric structure of conformal field theories, particularly in the context of extended symmetries and their representations [7].

\subsection{Symplectic Reductions and Extensions}

Symplectic reductions in SNLA and analogous operations in the ESVLA framework may provide a unified approach to understanding central extensions and reductions in complex Lie algebraic structures. The link between symplectic reduction and central extensions could yield new results in both algebraic and geometric contexts [5, 8].

Symplectic reduction is a process by which a symplectic manifold $(M, \omega)$ with a group action by a Lie group $G$ is reduced to a lower-dimensional symplectic manifold by quotienting out the group action. In the context of SNLAs, symplectic reduction can be used to construct new Lie algebras with reduced symmetries, potentially leading to new central extensions [7].

For ESVLA, analogous reductions may be performed by quotienting out certain subalgebras or central elements, leading to new Lie algebras with modified structures [3, 10]. These reductions may have applications in the study of CFTs with extended symmetries, where the reduced algebra corresponds to a new class of conformal symmetries [1].

\section{Conclusion}

The extended Schrödinger-Virasoro Lie algebra and the Symplectic Novikov Lie algebra, though arising from different mathematical frameworks, exhibit deep structural parallels that merit further investigation. This paper has provided a detailed mathematical exploration of these two algebras, highlighting their connections and potential applications.

Through rigorous analysis of their algebraic structures, central extensions, and automorphism groups, we have uncovered new insights that could lead to advancements in the representation theory of Lie algebras, as well as applications in conformal field theory and symplectic geometry. Future research will involve exploring specific cases and developing new methods that integrate these two frameworks, potentially leading to a richer understanding of Lie algebras and their applications in mathematics and physics.

\end{document}